\documentclass[aps,amssymb,floatfix,prd,amsmath,preprintnumbers]{revtex4}
\usepackage{epstopdf}
\usepackage{float}
\usepackage{capt-of}
\usepackage{graphicx}  
\usepackage{dcolumn}   
\usepackage{hyperref}
\usepackage{color}
\usepackage{url}
\usepackage{bm}
\begin{document}

\title{\bf Cosmological model with time varying deceleration parameter in $F(R,G)$ gravity}


\author{Santosh V. Lohakare\footnote{Department of Mathematics, Birla Institute of Technology and Science-Pilani, Hyderabad Campus, Hyderabad-500078, India, E-mail: lohakaresv@gmail.com}, S.K. Tripathy \footnote{Department of Physics, Indira Gandhi Institute of Technology, Sarang, Dhenkanal, Odisha-759146, India, E-mail: tripathy\_sunil@rediffmail.com}, B. Mishra\footnote{Department of Mathematics, Birla Institute of Technology and Science-Pilani, Hyderabad Campus, Hyderabad-500078, India, E-mail: bivu@hyderabad.bits-pilani.ac.in}, }

\affiliation { }

\begin{abstract}
\textbf{Abstract}: In this paper, we study the dynamical behaviour of the Universe in the $F(R,G)$ theory of gravity, where $R$ and $G$ respectively denote the Ricci scalar and Gauss-Bonnet invariant. Our wide analysis encompasses the energy conditions, cosmographic parameters, $Om(z)$ diagnostic, stability and the viability of reconstructing the referred model through a scalar field formalism.  The model obtained here shows the quintessence like behaviour at late times.  
\end{abstract} 

\maketitle

\section{Introduction} 

According to cosmological observations \cite{astier/2012,bengaly/2020}, the Universe is currently passing through an accelerated expansion phase. Generally, an entity named {\it dark energy} is said to cause such a counterintuitive anti-gravitational feature \cite{copeland/2006,frieman/2008,li/2011}. Most popular dark energy models advocate the presence of the cosmological constant in Einstein's field equations of General Relativity (GR) for a proper explanation of the observational data concerning the late time cosmic speed up issue \cite{riess/1998,peebles/2003,carroll/2001}. The cosmological constant may be thought of as the vacuum quantum energy in Particle Physics, whose estimated value \cite{weinberg/1989} differs more than one hundred orders of magnitude from the observed one \cite{riess/1998}. This characterizes the so-called {\it cosmological constant problem} \cite{weinberg/1989,martin/2012}.\\

The cosmological constant problem has been the main reason for which theoretical physicists consider extensions of GR. These extended gravity theories are obtained by simply substituting the Ricci scalar $R$ - in the Einstein-Hilbert action - by a generic function of $R$ itself and/or some other scalars. A replacement of $R$ by a more general function $F(R)$ in the Einstein-Hilbert action leads to the $F(R)$ gravity \cite{sotiriou/2010,de_felice/2010}. The $F(R)$ gravity applications are multitudinous. To quote some of them, Nojiri and Odintsov have obtained a unified cosmic history in \cite{nojiri/2011}, while the cosmological perturbations were investigated in \cite{matsumoto/2013,carloni/2008}, respectively by Matsumoto, and Carloni and collaborators.  Also, $F(R)$ theory of gravity has been extensively used in investigating the solutions of the wormhole geometry \cite{Samanta19,Samanta19a,Godani20}. Several cosmological solutions are shown as stable  in the literature for the $F(R)$ gravity theory \cite{Bohmer07,Shah19,Pretel20}.\\

However it is well-known that $F(R)$ gravity is plagued by some shortcomings. For instance, Frolov has pointed out a curvature singularity problem appearing on the non-linear level \cite{frolov/2008}. Kobayashi and Maeda argued that relativistic stars cannot be present in $F(R)$ theories \cite{kobayashi/2008}. This has been revisited in \cite{kobayashi/2009,goswami/2014}. Furthermore, Solar System regime constraints obtained from GR classical tests seem to rule out most of the $F(R)$ models proposed so far \cite{chiba/2003,chiba/2007,olmo/2007}.\\

With the purpose of circumventing these shortcomings, the $F(R)$ gravity has been extended through the consideration of further scalars in the referred Einstein-Hilbert action. In this regard, the $F(R,G)$ gravity \cite{de_laurentis/2015,wu/2015,santos_da_costa/2018,farasat_shamir/2018,odintsov/2019,kumar_sanyal/2020,singh/2021}, with $G$ being the Gauss-Bonnet scalar, arises as an optimistic alternative. For instance, a double inflationary scenario naturally emerges in the $F(R,G)$ gravity according to \cite{de_laurentis/2015}. The stability of cosmological solutions in $F(R,G)$ gravity is discussed in \cite{de_la_cruz-dombriz/2012}. A study of linear metric perturbations around a spherically symmetric static space-time in $F(R,G)$ theories can be seen in \cite{de_felice/2011}. In \cite{elizalde/2010} it is shown that $F(R,G)$ cosmology naturally leads to an effective cosmological constant, quintessence or phantom cosmic acceleration, not without describing the transition from a decelerated stage of the Universe expansion. In the present article we wish to construct a viable cosmological model in the framework of the $F(R,G)$ gravity and to analyse the energy conditions of the constructed model.\\

The energy conditions are mathematical inequalities that basically state that energy density cannot be negative. They are written in terms of the energy-momentum tensor as we are going to show below in Section IV. For now it is worth mentioning that in \cite{lima/2008} a picture of energy conditions fulfilment and violation in the light of type Ia supernovae observational data was provided. Seminal applications of energy conditions can be seen in \cite{morris/1988,ford/1995,tipler/1978}. In extended gravity, the energy conditions have also shown great applicability and commendable results as one can check \cite{mehdizadeh/2015,pennalima/2018}. Particularly, some viability bounds were put to $F(R,G)$ gravity from the energy conditions in \cite{singh/2021}. \\

Here we will also analyse the cosmography, $Om(z)$ diagnostic and the dynamical stability under linear homogeneous perturbations, and finally the viability of the model through its scalar field reconstruction is presented.\\

Our article is organized as follows: in Section II we present the $F(R,G)$ gravity and cosmology basics. The material solutions of a particular $F(R,G)$ model are presented in Section III in terms of redshift. The behaviour of the model with the  energy conditions, geometrical parameters, $Om(z)$ diagnostic, scalar reconstruction are done in Section IV. The concluding remarks are given in Section V.

\section{Basic Formalism of $F(R,G)$ Gravity and Cosmology}

An interesting modification to the Einstein theory of gravity is
the $F(R, G)$ gravity \cite{de_laurentis/2015,wu/2015,santos_da_costa/2018,farasat_shamir/2018,odintsov/2019,kumar_sanyal/2020,singh/2021}. The most general action for this gravity is given by
\begin{equation}\label{1}
S=\int \sqrt{-g}\left[\frac{1}{2k^2}F(R,G)+\mathcal{L}_m\right] d^{4}x
\end{equation}
where $g$ is the metric determinant, $\mathcal{L}_m$ describes the matter Lagrangian, $k^2=8\pi G_N$, $G_N$ is the Newtonian gravitational constant and the speed of light $c$ is taken as 1. The Gauss-Bonnet invariant is described as
\begin{equation}\label{2}
G \equiv R^2-4R^{\mu \nu} R_{\mu \nu}+R^{\mu \nu \alpha \beta}R_{\mu \nu \alpha \beta},
\end{equation}
with $R^{\mu\nu}$ being the Ricci tensor and $R^{\mu\nu\alpha\beta}$ the Riemann tensor. In differential geometry, $G$ can be described as
\begin{equation}\label{3}
\int_{\mathcal{M}} G d^{n}x=\chi (\mathcal{M}),
\end{equation}
where $\chi(\mathcal{M})$ are the Euler characteristics of the manifold $\mathcal{M}$ in $n$ dimensions. For $n=4$, $\chi(\mathcal{M})=0$, it can be considered as surface term that does not affect the dynamics. By varying the action \eqref{1} with respect to the metric tensor $g_{\mu \nu}$, the field equations of the $F(R,G)$ gravity can be expressed as
\begin{eqnarray} \label{4}
\nonumber F_R{G}_{\mu\nu}&=&k^2{T}^{m}_{\mu\nu}+\frac{1}{2}g_{\mu\nu}[F(R,G)-RF_{R}]+\nabla_{\mu}\nabla_{\nu} F_{R}-g_{\mu\nu} \Box F_{R}\\\nonumber&+&F_{G}({-2R}{R_{\mu\nu}}+4R_{\mu k}R^{k}_{\nu}-2R^{klm}_{\mu}R_{\nu k l m}+4g^{kl} g^{mn} R_{\mu k \nu m} R_{ln})\\\nonumber&+&2(\nabla_{\mu}\nabla_{\nu}F_{G})R-2g_{\mu \nu}(\Box F_{G})R+4(\Box F_{G})R_{\mu\nu}-4(\nabla_{k} \nabla_{\mu} F_{G})R^{k}_{\nu}\\&-&4(\nabla_{k} \nabla_{\nu} F_{G})R^{k}_{\mu}+4g_{\mu \nu}(\nabla_{k} \nabla_{l} F_{G})R^{kl}-4(\nabla_{l} \nabla_{n} F_{G})g^{kl}g^{mn}R_{\mu k \nu m}
\end{eqnarray}
where $G_{\mu \nu}$ is the Einstein tensor, $\nabla_{\mu}$ is the covariant derivative operator associated with $g_{\mu \nu}$, $\Box \equiv g^{\mu \nu}\nabla_{\mu}\nabla_{\nu}$ is the covariant d'Alembert operator and ${T}_{\mu\nu}$ is the energy-momentum tensor. We have also defined the following quantities
\begin{equation} \label{5}
F_{R}\equiv \frac{\partial F(R,G)}{\partial R},\hspace{1cm} F_{G}\equiv \frac{\partial F(R,G)}{\partial G},
\end{equation}
while the energy-momentum tensor is written as
\begin{equation}\label{6}
T_{\mu\nu}=(\rho+p)u_{\mu}u_{\nu}+pg_{\mu\nu},
\end{equation}
where $\rho$ denotes the matter-energy density and $p$ is the isotropic pressure measured by the observer $u_{\mu}$.

In the following we consider the spatially flat Friedmann-Lem\^aitre-Robertson-Walker metric with line element,

\begin{equation} \label{7}
ds^{2}=-dt^{2}+a^{2}(t)(dx^{2}+dy^{2}+dz^{2}),
\end{equation}
where $a(t)$ is the scale factor of the Universe, such that the Hubble parameter $H\equiv\frac{\dot{a}}{a}$, with the over dot indicating derivative with respect to cosmic time $t$. Then, $R$ and $G$ become,

\begin{equation} \label{8}
R=6(\dot{H}+2H^{2}), \hspace{2cm} G=24H^{2}(\dot{H}+H^{2}).
\end{equation}

By substituting \eqref{6} and \eqref{7} into the gravitational field equations \eqref{4} we obtain 

\begin{eqnarray} 
3H^{2}&=&\frac{\kappa^{2}}{F_{R}} \rho+\frac{1}{2}\left[R+G\frac{F_{G}}{F_{R}} - \frac{F(R,G)}{F_{R}}\right]-3H \frac{\dot{F_{R}}}{F_{R}}-12H^{3}\frac{\dot{F_{G}}}{F_{R}},\label{9}\\
2\dot{H}+3H^{2} &=& -\frac{\kappa^{2}}{F_{R}} p +\frac{1}{2}\left[R+G\frac{F_{G}}{F_{R}} - \frac{F(R,G)}{F_{R}}\right]-2H\frac{\dot{F_{R}}}{F_{R}}-\frac{\ddot{F_{R}}}{F_{R}}-4H^{2}\frac{\ddot{F_{G}}}{F_{R}}-8H\left(\dot{H}+H^2\right)\frac{\dot{F_{G}}}{F_{R}}.\label{10}
\end{eqnarray}
Eqns. \eqref{9} and  \eqref{10} are expressed in terms of the Hubble parameter and the functional $F(R,G)$. To obtain the evolutionary behaviour of the matter pressure and energy density, we need a functional form for $F(R,G)$ and the Hubble parameter, which we have discussed in the following section.   
\section{Material Solutions}

In the present section, we extend our analysis by considering a specific form for the $F(R, G)$ function. We take
\begin{equation} \label{11}
F(R,G)=R+\alpha R^{2}+\beta G^{2}
\end{equation}
where $\alpha$ and $\beta$ are constants. This particular functional form was used, for instance, in \cite{de_laurentis/2015}, in which a double inflationary scenario has naturally emerged. By using the form of $F(R,G)$ as in \eqref{11}, Eqs.\eqref{9} and \eqref{10} take the form

\begin{eqnarray}
\rho&=&\frac{1}{\kappa^{2}}(3H^2+108\alpha \dot{H} H^2-18\alpha \dot{H^2}-288\beta H^8+1728\beta \dot{H} H^6+864\beta \dot{H^2}H^4+36\alpha H\ddot{H}+576\beta\ddot{H}H^5),\label{12}\\ 
\nonumber p&=&\frac{1}{\kappa^{2}}(-2\dot{H}-3H^2-54\alpha \dot{H^2}-108\alpha \dot{H} H^2+288\beta H^8-960\beta \dot{H} H^6-4320\beta \dot{H^2}H^4-72\alpha H\ddot{H}-12\alpha \dddot{H}\\&-&1152\beta \ddot{H}H^5-1152\beta H^2 \dot{H^3}-1536\beta \dot{H} \ddot{H} H^3-192\beta \dddot{H}H^4).\label{13}
\end{eqnarray}

\begin{figure}[!htp]
\centering
\includegraphics[scale=0.50]{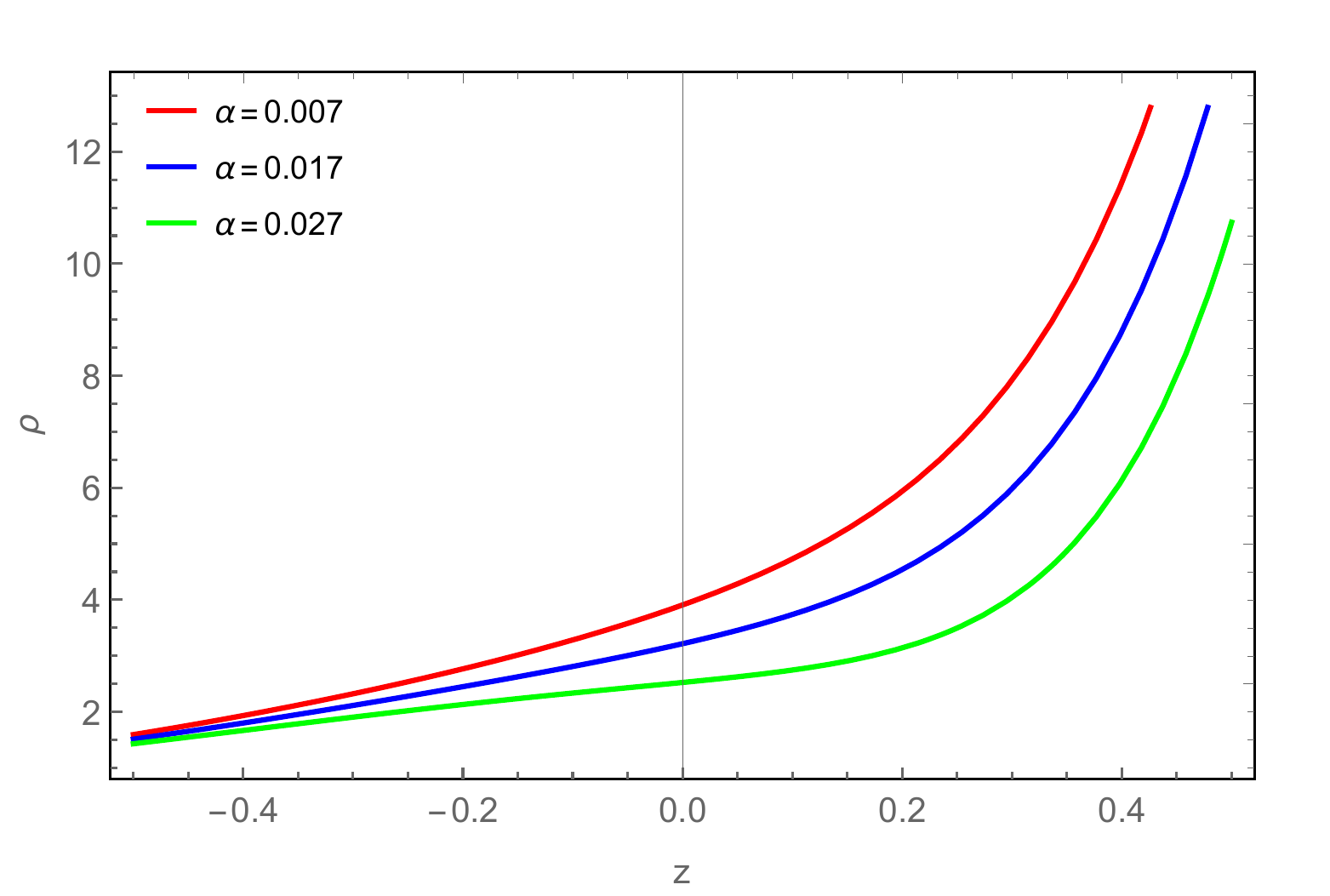}
\includegraphics[scale=0.50]{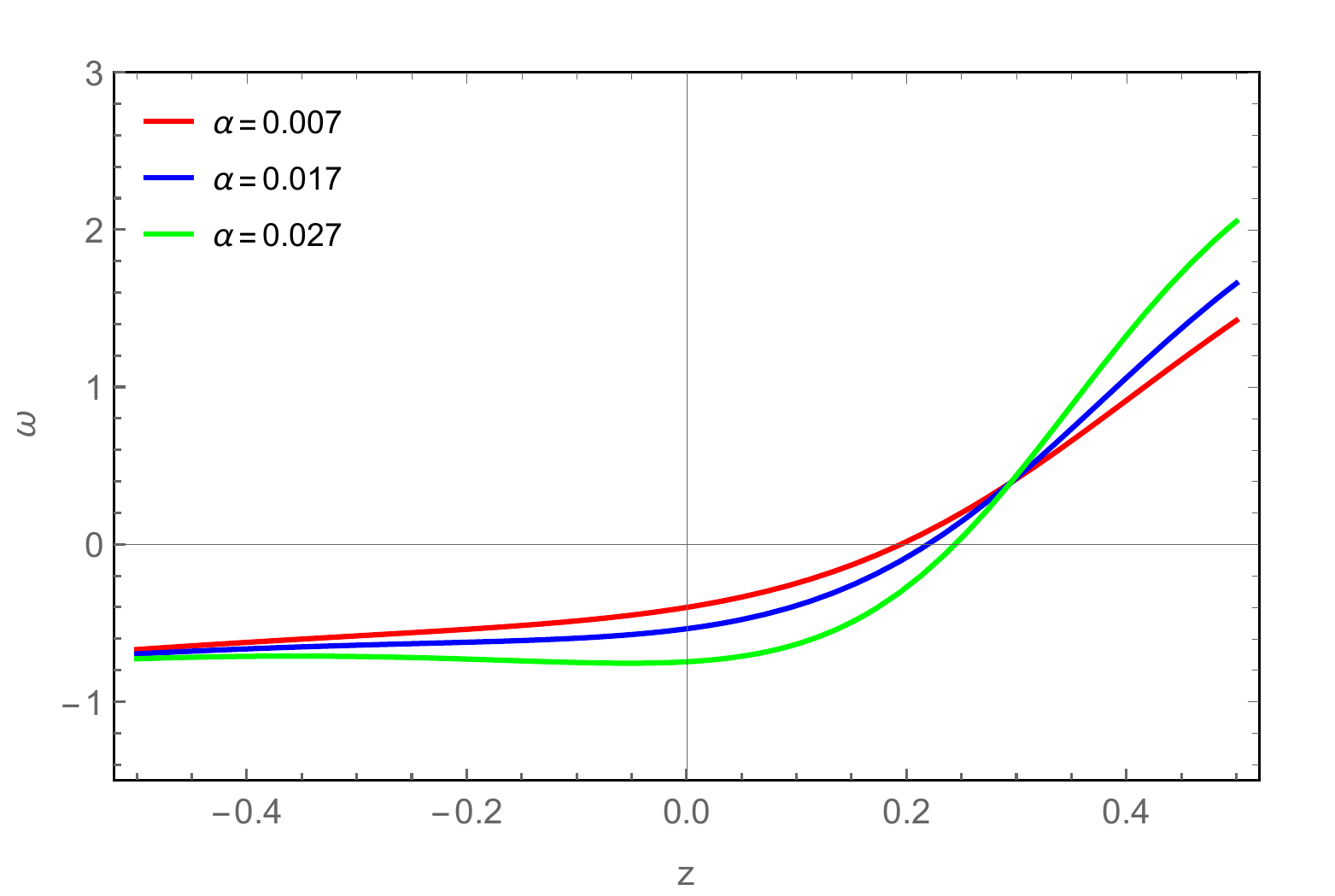}
\caption{Graphical behaviour of energy density (left panel) and EoS parameter (right panel) versus redshift $z$ with the representative value of the parameters $\eta=0.32$, $\nu=0.614$ and $\beta=0.00018.$ }
\label{Fig1}
\end{figure}

We have expressed the field equations of $F(R,G)$ gravity in the form of Hubble parameter. The dynamical parameters can be determined either with a relationship between the matter field or with an assumed form of the Hubble parameter. We have preferred here to frame the cosmological model with an assumed form of scale factor. Also, to frame a cosmological model of the Universe we need to obtain the pressure and energy density with respect to the cosmic time or redshift. In order to handle Eqs.\eqref{12} and \eqref{13}, which are highly non-linear, we assume the hybrid scale factor (HSF), $a(t)=e^{{\eta}{t}} t^{\nu}$, such that $H=\eta+\frac{\nu}{t}$, where $\eta$ and $\nu$ are the model parameters and can be constrained in the ranges $\eta>0$ and $0<\nu<1$ \cite{Mishra15, SKT2020, SKT2021,SKT21a,Pati2021}. The reason behind the choice of such a scale factor is that it simulates a transition from a decelerated expanding Universe to an accelerated one. This can be substituted in Eqs.\eqref{12}-\eqref{13} to obtain the expressions for the pressure and energy density in terms of cosmic time. We obtain lengthy expressions for the energy density and pressure and therefore we opt to present them in graphical form. Along with the energy density, we have also presented the equation of state parameter (EoS), $\omega=\frac{p}{\rho}$ graphically. In addition we prefer to present the graphs in terms of redshift $z$ with the use of the expression $\frac{1}{a}=1+z$ for better analysis.  The energy density $\rho$ and the EoS parameter $\omega=p/\rho$ behaviours can be seen in  FIG.\ref{Fig1}. To note, in theoretical sense, the behaviour of the EoS parameter is utmost essential to address the late time cosmic acceleration of the Universe. Then with the already obtained value of the scale factor parameter, we will analyse the EoS parameter to find the present and future scenario of the Universe. \\

\begin{table}
\caption{Model parameters.}
\centering
\begin{tabular}{c|c}
\hline
\hline
Parameters  & Constrained values  \\
\hline
$\eta$     & 0.32  \\
\hline
$\nu$      &0.614   \\
\hline
$\beta$	 &0.00018\\
\hline
$\alpha$   &0.007 \\
		 &0.017\\
		 &0.027\\
\hline
\end{tabular}
\end{table}

The evolutionary behaviour of matter pressure and energy density depend on the $F(R,G)$ parameters $\alpha$, $\beta$ and scale factor parameters $\eta$ and $\nu$. The scale factor parameters $\eta$ and $\nu$ are chosen similar to the works \cite{Tripathy20a,SKT2020,Mishra21} to obtain an appropriate behaviour of the geometrical parameters. Now, the parameters of the assumed form of $F(R,G)$ gravity, $\alpha$ and $\beta$ are fixed to obtain the positive energy density and negative pressure to address the late time cosmic expansion phenomena. We have examined several possible combinations of the value of $\alpha$ and $\beta$ and observed that with the lower value of $\beta$, the model shows appropriate accelerating behaviour. So, a small contribution from the Gauss-Bonnet invariant suffices for the late time cosmic acceleration. In Table-I, we have presented the model parameters considered in the present work. FIG. \ref{Fig1}(left panel) shows that the energy density decreasing slowly from a higher positive value and approaching to a small value at the late time. Another observation is that with the increase in the parametric value $F(R,G)$ parameter $\alpha$, the evolution of energy density is starting from a lower value and for the other parameter $\beta$, no significant changes have been noticed. However, throughout the evolution, the energy density entirely remains in the positive region. The EoS parameter that shows us the behaviour of the accelerating Universe is found to be evolve from a early positive value and approaching $\thickapprox -0.7$ at late times. The curves of the EoS parameter are observed to decrease faster for higher value of $\alpha$. However, at $z\thickapprox 0.3$, the curves intersect. We do not observe any substantial changes on the evolutionary behaviour for a variation in the values of $\beta$.  We can conclude that the present model in the framework of  $F(R,G)$ gravity and hybrid scale factor favours a quintessence like evolutionary phase. 

\section{Analysis of the Model}
\subsection{Energy conditions in $F(R,G$) gravity}
The energy conditions are basically the boundary conditions in order to keep the energy density positive \cite{Hawking73,Poisson04}. These do not correspond to the physical reality and the most recent example is the violation of strong energy condition with the observable effects of dark energy. Energy conditions provide additional constraints to the cosmological models \cite{Carroll04}. The energy conditions assign the fundamental causal and geodesic structure of the space-time and the extended theory of gravity needs to confront with the energy conditions, in this case with the $F(R,G)$ gravity. So, any extended theory of gravity, which is an extension of Einstein's gravity can be recasted in such a manner that it can be dealt with the standard energy conditions \cite{Nojiri11,Capozziello11}. The energy conditions are, Null Energy Condition (NEC): $\rho+p\geq 0 $, Weak Energy Condition (WEC): $\rho+p \geq 0$, $\rho\geq0$, Strong Energy Condition (SEC): $\rho+3p \geq 0$ and Dominant Energy Condition (DEC): $\rho-p \geq 0$. With the help of Eqs. \eqref{12} and \eqref{13}, we can obtain the energy conditions with respect to the cosmic time as,

\begin{eqnarray}
\rho+p &=&-72\alpha \dot{H}^2-3456 \beta H^4 -36 \alpha H\ddot{H}-576\beta H^5 \ddot{H}-12\alpha H^3 \nonumber \\
&-&192 \beta H^7 -1536\beta H^3 \dot{H} \ddot{H}+768 \beta H^6 \dot{H}-2\dot{H}-1152 \beta H^2 \dot{H}^3 \geq0 \label{14}
\end{eqnarray}

\begin{eqnarray}
\rho+p &=&-72\alpha \dot{H}^2-3456 \beta H^4 -36 \alpha H\ddot{H}-576\beta H^5 \ddot{H}-12\alpha H^3 \nonumber \\
&-&192 \beta H^7 -1536\beta H^3 \dot{H} \ddot{H}+768 \beta H^6 \dot{H}-2\dot{H}-1152 \beta H^2 \dot{H}^3 \geq0; ~~~\rho\geq0 
\end{eqnarray} \label{15}

\begin{eqnarray}
\rho+3p &=&-216 \alpha \dot{H}H^2+4608 \beta  H^3 \dot{H} \ddot{H}+1152\beta \dot{H}  H^6+ 6\dot{H}- 30 \alpha \dot{H}^2-2016 \beta  H^4\dot{H}^2 \nonumber \\
&-&30 \alpha H \ddot{H} -480 \beta H^5 \ddot{H}-6 \alpha H^3-96\beta H^7 -3456\beta  H^2 \dot{H}^3+ 576 \beta H^8 -6H^2\geq0 \label{16}
\end{eqnarray}

\begin{eqnarray}
\rho-p&=& 216 \alpha H^2 \dot{H}+1536 \beta  H^3 \dot{H} \ddot{H}+2688 \beta \dot{H}  H^6+2\dot{H}+ 36 \alpha \dot{H}^2 +5184 \beta  H^4 \dot{H}^2 \nonumber\\
&+& 108\alpha H \ddot{H}+1728 \beta H^5 \ddot{H}+12 \alpha H^3+192 \beta H^7-576 \beta  H^8+6H^2+1152 \beta  H^2 \dot{H}^3\geq0 \label{17}
\end{eqnarray}

\begin{figure}[!htp]
\centering
\includegraphics[scale=0.50]{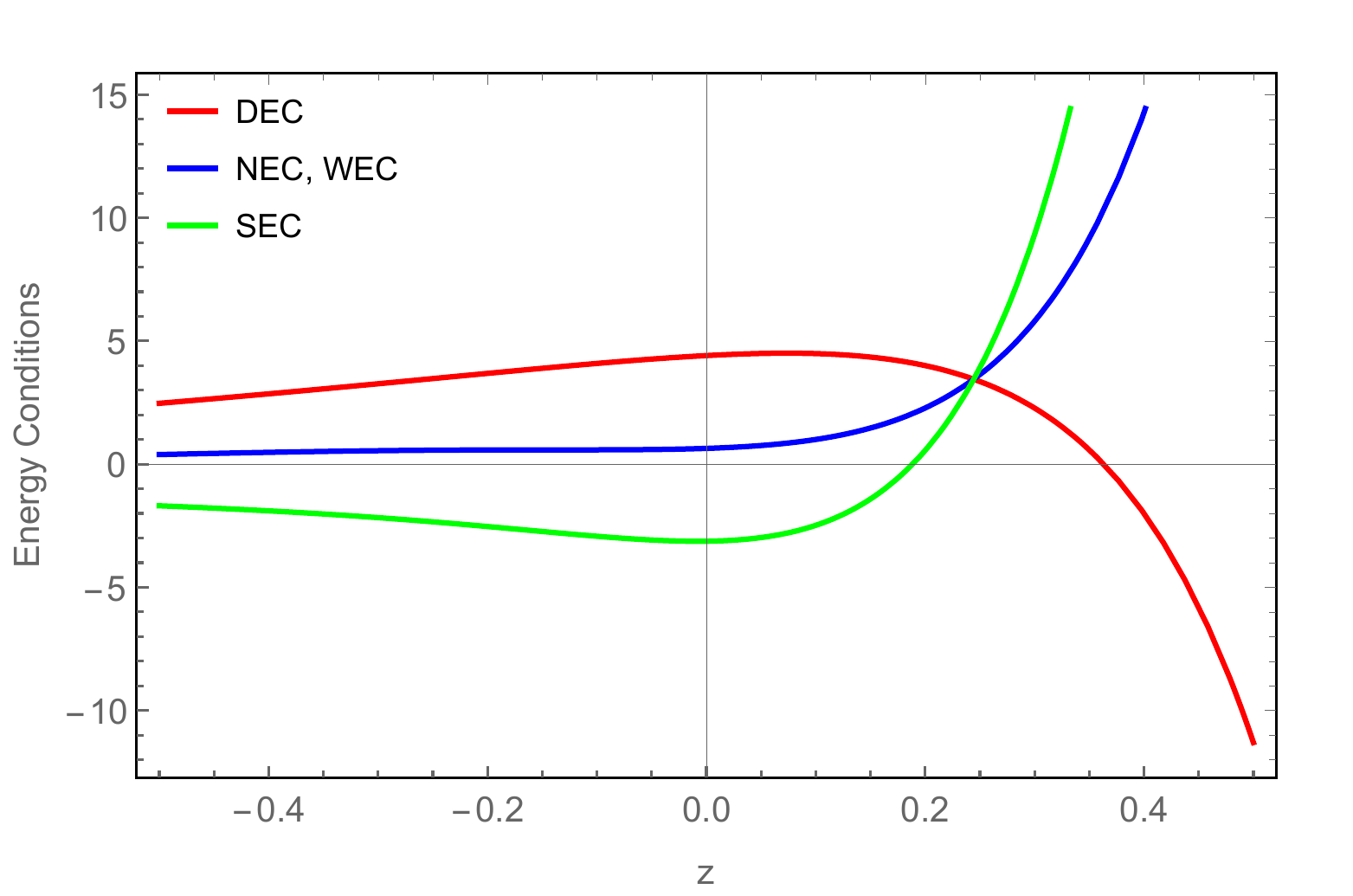}
\includegraphics[scale=0.50]{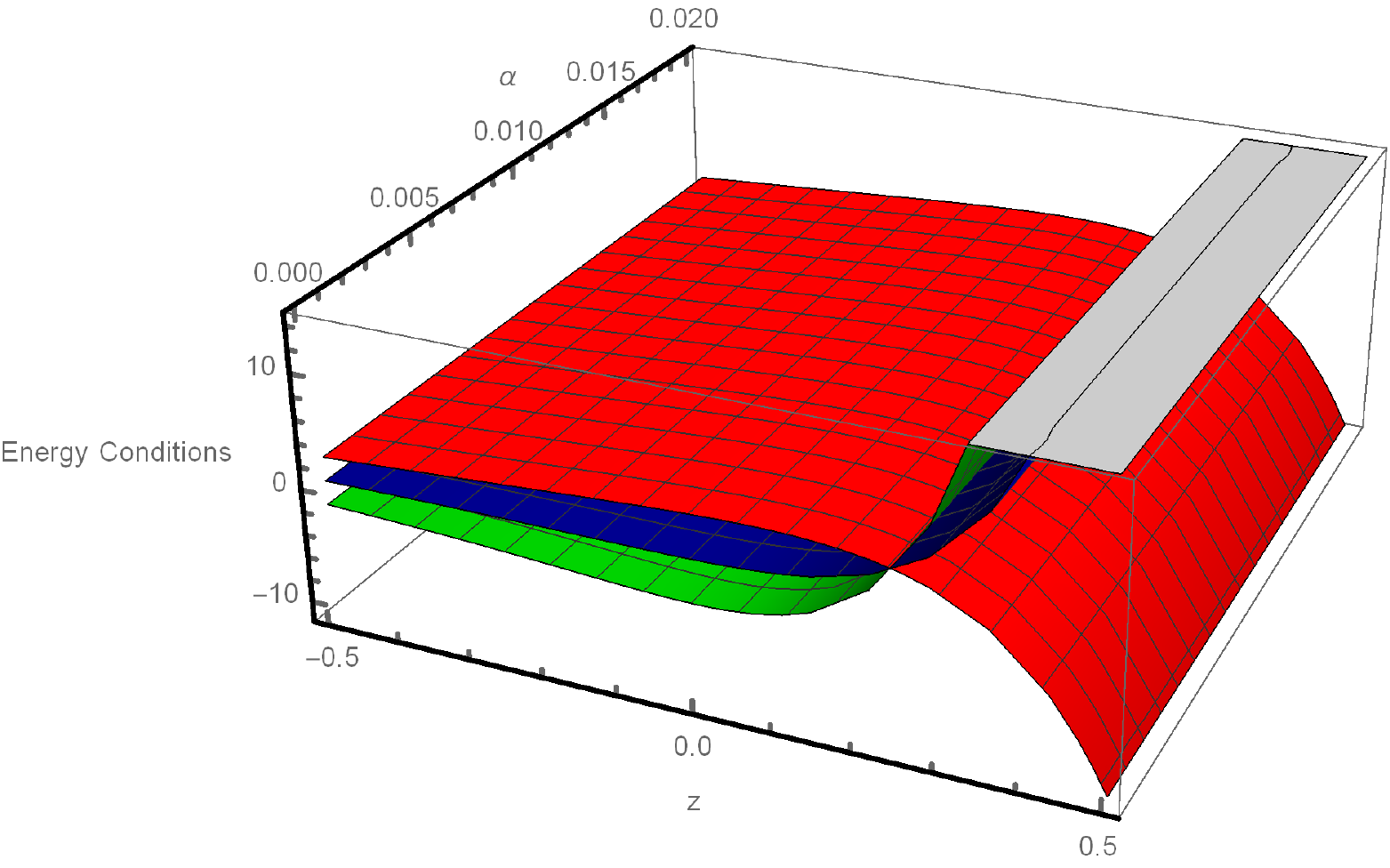}
\caption{Graphical behaviour of energy conditions versus redshift $z$(2D, left panel),  (3D, right panel), $\eta=0.32$, $\nu=0.614$, $\alpha=0.027$, $\beta=0.00018.$}
\label{Fig2}
\end{figure}

Another reason to fix the representative value of the model and scale factor parameters above is that we wish to compel our model in such a manner that the WEC condition satisfied at least at the late phase. FIG. (\ref{Fig2}) shows the WEC remains positive from an early time ($z\thickapprox0.35$) till the late phase. Since our model is showing quintessence behaviour, we may expect the satisfaction of DEC and NEC at least at the late phase of the evolution. At the same time, the SEC starts violating from ($z\thickapprox0.2$) and was satisfying before that. In fact, when the cosmic dynamics is fixed up by a derived or assumed Hubble rate, the detailed analysis on these energy condition can be obtained.

\subsection{Geometric Parameters and $Om(z)$ Diagnostic }
There are two fundamental characteristics of the evolution of the Universe: (i) it has been extracted directly from the space-time metric; the kinematic approach and (ii) the dependency on the properties of the field that fill the Universe; the dynamic approach. The kinematic characteristic is universal, which is convenient to describe the expansion of the universe, whereas the dynamic characteristic is model dependent. Here we shall follow the kinematic approach. We have considered the FLRW space-time, which is homogeneous and isotropic, so the evolution of the Universe  can be described by the scale factor $a(t)$. The scale factor $a(t)$ and the co-moving coordinate $r$ can be related to the physical Euler coordinate $R(t)$ as, $a(t)=R(t)/r$. On differentiation with respect to time, we can obtain the Hubble law in the form, $V=HR$, where $V=\dot{R}=dR/dt$ and $H=\frac{\dot{a}}{a}$ be the Hubble parameter. The expansion of the Universe, as determined by the Hubble parameter, depends on time and the measure of this dependency is the deceleration parameter. We shall define here the Taylor series expansion of the scale factor in the neighbourhood of the current time $t_0$,

\begin{equation}
a(t)= a(t_0)+ \dot{a}(t_0)[t-t_0]+\frac{1}{2}\ddot{a}[t-t_0]^2+...\label{18}
\end{equation}

This scheme of the description of the universe is known as cosmography \cite{Weinberg72} that is based on the cosmological principle. As per the principle of cosmology, the scale factor act as the degree of freedom in governing the Universe. The parameter set Hubble parameter ($H$), deceleration parameter ($q$), jerk parameter ($j$), snap parameter ($s$), lerk parameter ($l$),.... represents the alphabet of the cosmography. This set can be obtained upto the fifth derivative of the scale factor, where the first derivative is the Hubble parameter and so on as in following,\\

Deceleration parameter, $q(t)\equiv-\frac{1}{a} \frac{{d^2}a}{dt^2} \left[\frac{1}{a} \frac{da}{dt}\right]^{-2}=-1+\frac{\nu }{(\nu +\eta  t)^2}$,\\

Jerk parameter, $j(t)\equiv\frac{1}{a} \frac{{d^3}a}{dt^3} \left[\frac{1}{a} \frac{da}{dt}\right]^{-3}=\frac{2 \nu +(\nu +\eta  t) \left[(\nu +\eta  t)^2-3 \nu \right]}{(\nu +\eta  t)^3}$,\\

Snap parameter, $s(t)\equiv\frac{1}{a} \frac{{d^4}a}{dt^4} \left[\frac{1}{a} \frac{da}{dt}\right]^{-4}=\frac{(\nu -3) (\nu -2) (\nu -1) \nu +4 \eta ^3 \nu  t^3+6 \eta ^2 (\nu -1) \nu  t^2+\eta ^4 t^4+4 \eta  (\nu -2) (\nu -1) \nu  t}{(\nu +\eta  t)^4}$,\\

and Lerk parameter,\\
$l(t)\equiv\frac{1}{a} \frac{{d^5}a}{dt^5} \left[\frac{1}{a} \frac{da}{dt}\right]^{-5}=\frac{(\nu -4) (\nu -3) (\nu -2) (\nu -1) \nu +5 \eta ^4 \nu  t^4+10 \eta ^3 (\nu -1) \nu  t^3+10 \eta ^2 (\nu -2) (\nu -1) \nu  t^2+\eta ^5 t^5+5 \eta  (\nu -3) (\nu -2) (\nu -1) \nu  t}{(\nu +\eta  t)^5}$.\\

\begin{figure}[!htp]
\centering
\includegraphics[scale=0.50]{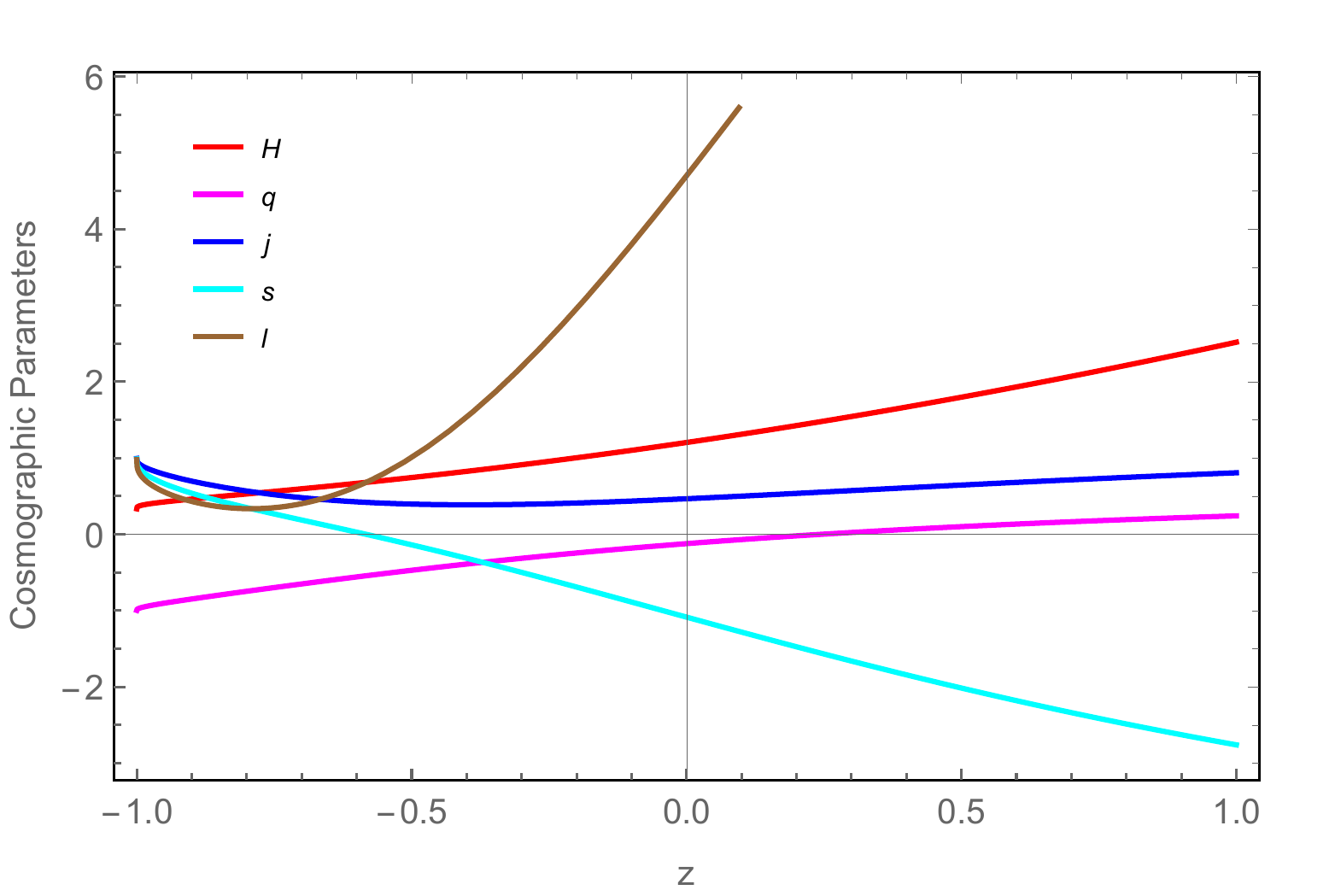}
\includegraphics[scale=0.50]{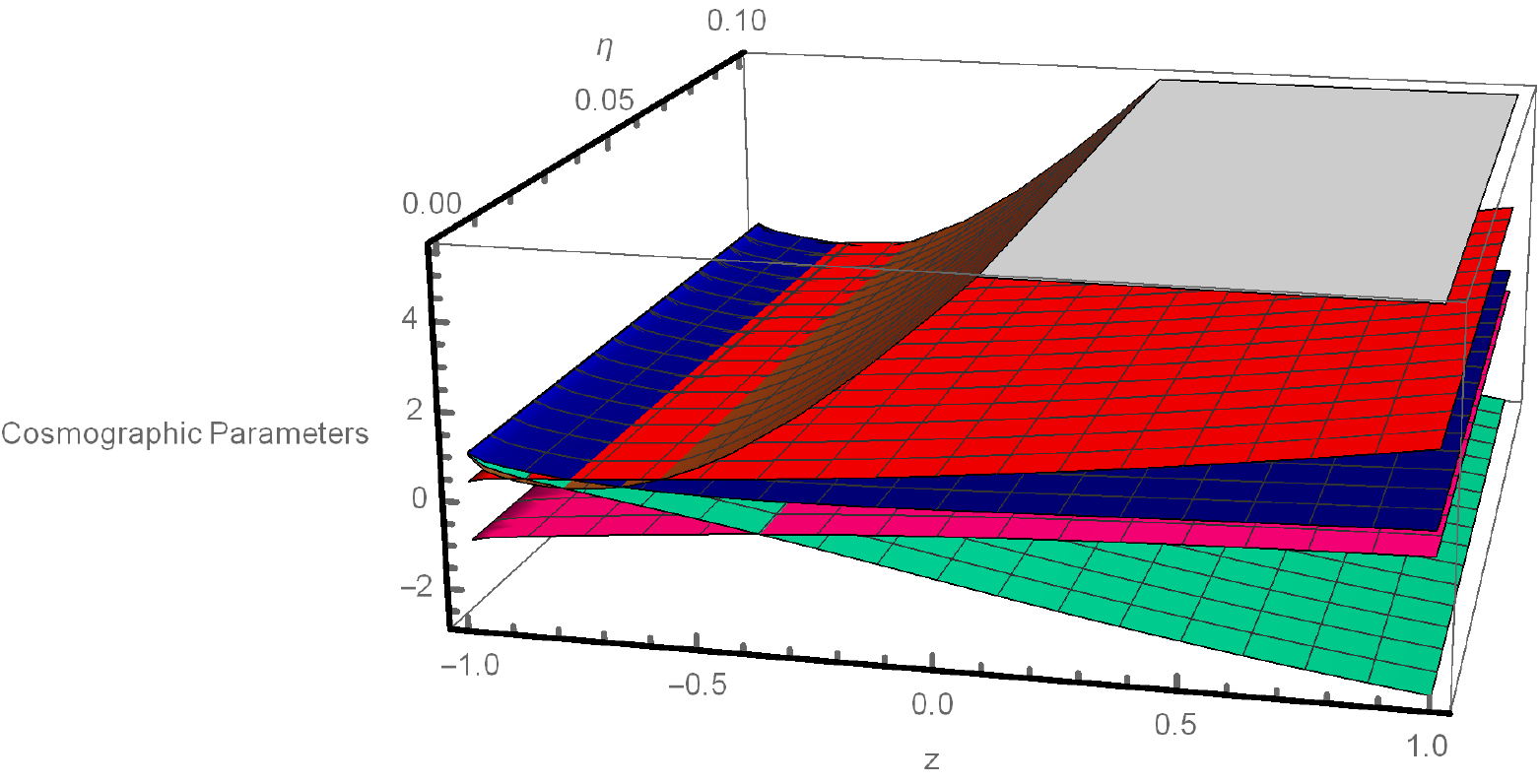}
\caption{Graphical behaviour of the cosmographic parameters (2D, left panel), (3D, right panel)versus redshift $z$ with the representative value of the parameters, $\eta=0.32$, $\nu=0.614.$}
\label{Fig3}
\end{figure}

The graphical behaviour of the cosmographic parameter set has been given in FIG. \ref{Fig3}. The Hubble parameter, $H=\eta+\frac{\nu}{t}$ becomes a constant ($\eta$) at late times. It remains entirely in the positive domain since the scale factor parameters are fixed with the positive value (Red curve). The deceleration parameter is showing the signature flipping behaviour and at $t\rightarrow 0$, $q\simeq-1+\frac{1}{\nu}$ whereas at $t\rightarrow 0$, $q\simeq-1$. So, the hybrid scale factor gives a deceleration parameter that assumes early positive and late time negative values. To mimic the present Universe with late time cosmic acceleration, there is a need of signature flipping behaviour of the deceleration parameter. This behaviour also favouring the recent $H_0$ tension findings that a transitive deceleration parameter fostering an early deceleration and late time acceleration is in accordance with the concordance $\Lambda$CDM model \cite{Riess16,Riess18,Aghanim20}. One can observe the similar feature in FIG.\ref{Fig3}(pink curve). The jerk parameter decreases from a higher positive value and remains entirely in the positive region (Blue curve) whereas the snap parameter is showing the transition behaviour from negative to positive value at late times. Since our model favours the quintessence behaviour therefore $j<1$ and $s>0$. Similar behaviour has been observed for both these geometrical parameters at least at the late times. The lerk parameter is decreasing very rapidly, then increases a little bit and settled at $\simeq 0.5$. To note except the deceleration parameter, all other geometrical parameters are lying in the range $(0,1)$ at late phase of the evolution. \\

\begin{figure}[!htp]
\centering
\includegraphics[scale=0.50]{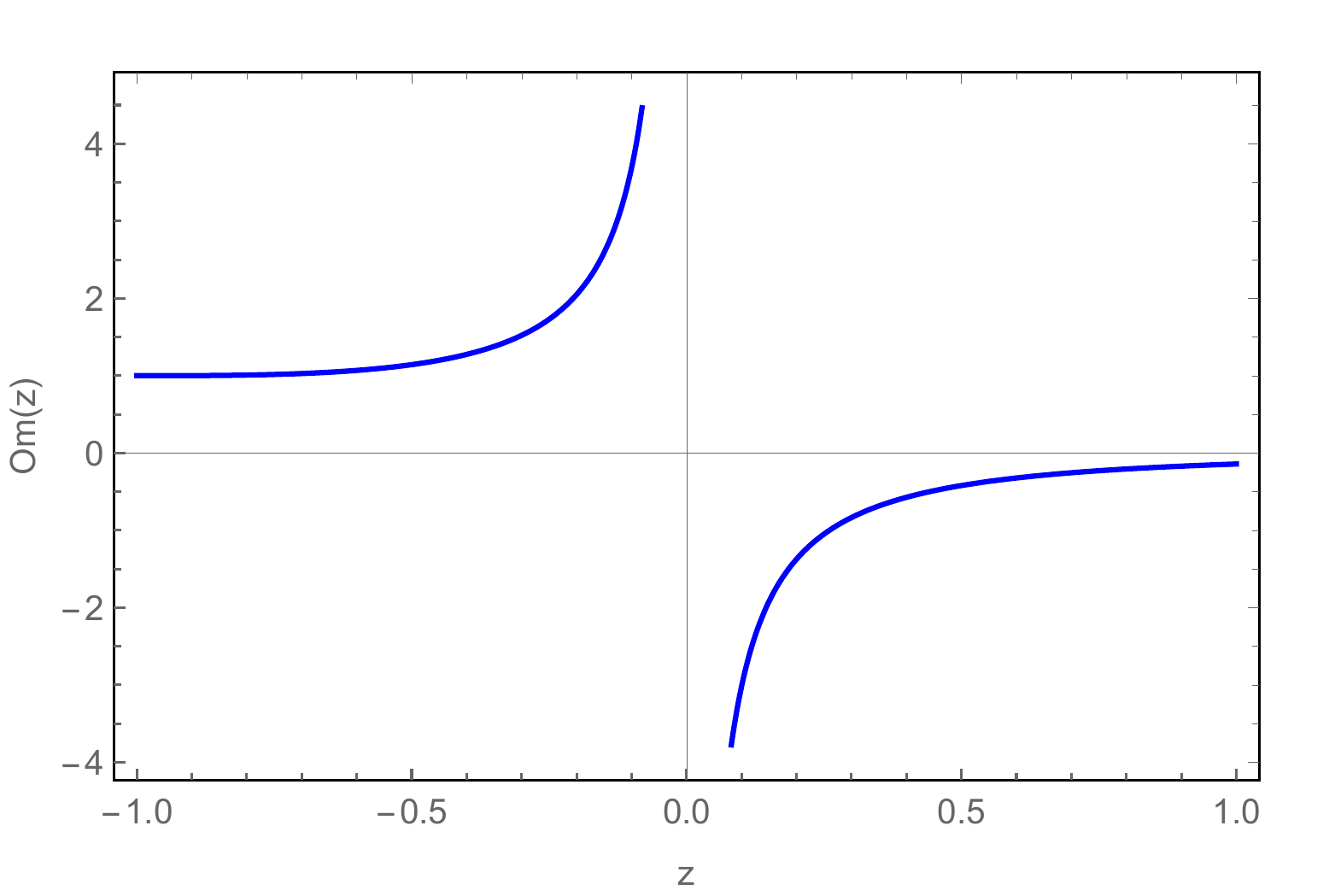}
\caption{Graphical behaviour of $Om(z)$ vs redshift $z$  with the representative value of the parameters, $\eta=0.32$, $\nu=0.614$. }
\label{Fig4}
\end{figure}

There are two important geometrical diagnostic approaches used in literature. They are the determination of the state finder pair $(j,s)$ in the $j-s$ plane and the $Om(z)$ diagnostics. These geometrical diagnostic approaches are useful tools to distinguish different dark energy models \citep{Alam03,Sahni08}. The behaviour of the state finder pair can be seen from FIG. \ref{Fig3}. The $Om(z)$  diagnostic provides a null test to the $\Lambda$CDM model \cite{Sahni08} and subsequently more evidences were gathered on its sensitiveness with the EoS parameter \cite{Ding15,Zheng16,Qi18}. When $Om(z)$ is constant with respect to the redshift, then the dark energy model would be in the form of cosmological constant. Also, the nature of the slope of $Om(z)$ distinguish the dark energy models as: positive slope of the evolving $Om(z)$ indicates phantom phase $\omega<-1$, and negative slope for quintessence region $\omega>-1$. Also the consistency test of $\Lambda$CDM model has been performed in the reconstructed $Om(z)$ by using the Gausssian processes with SN Ia and Hubble data set \cite{Yahya14,Qi16}. In this problem, we wish to investigate the behaviour of the model with the $Om(z)$ diagnostic which can be defined as,

\begin{equation}
Om(z)=\frac{E^2(z)-1}{(1+z)^3 -1}\label{19}
\end{equation}
where $E(z)=\frac{H(z)}{H_0}$ is the dimensionless parameter, here $H_0$ be the Hubble rate of the present epoch. The plot for $Om(z)$  with respect to redshift has been given in FIG. \ref{Fig4}. The $Om(z)$ parameter is showing a discrete behaviour. At an early time it remains entirely in the negative domain whereas at late times in the positive domain. At both times the parameter decreases from higher to lower value. At early time it decreases indefinitely and at late time it deceases gradually and settled near $1$.

\subsection{Scalar Field Reconstruction}

\begin{figure}[!htp]
\centering
\includegraphics[scale=0.50]{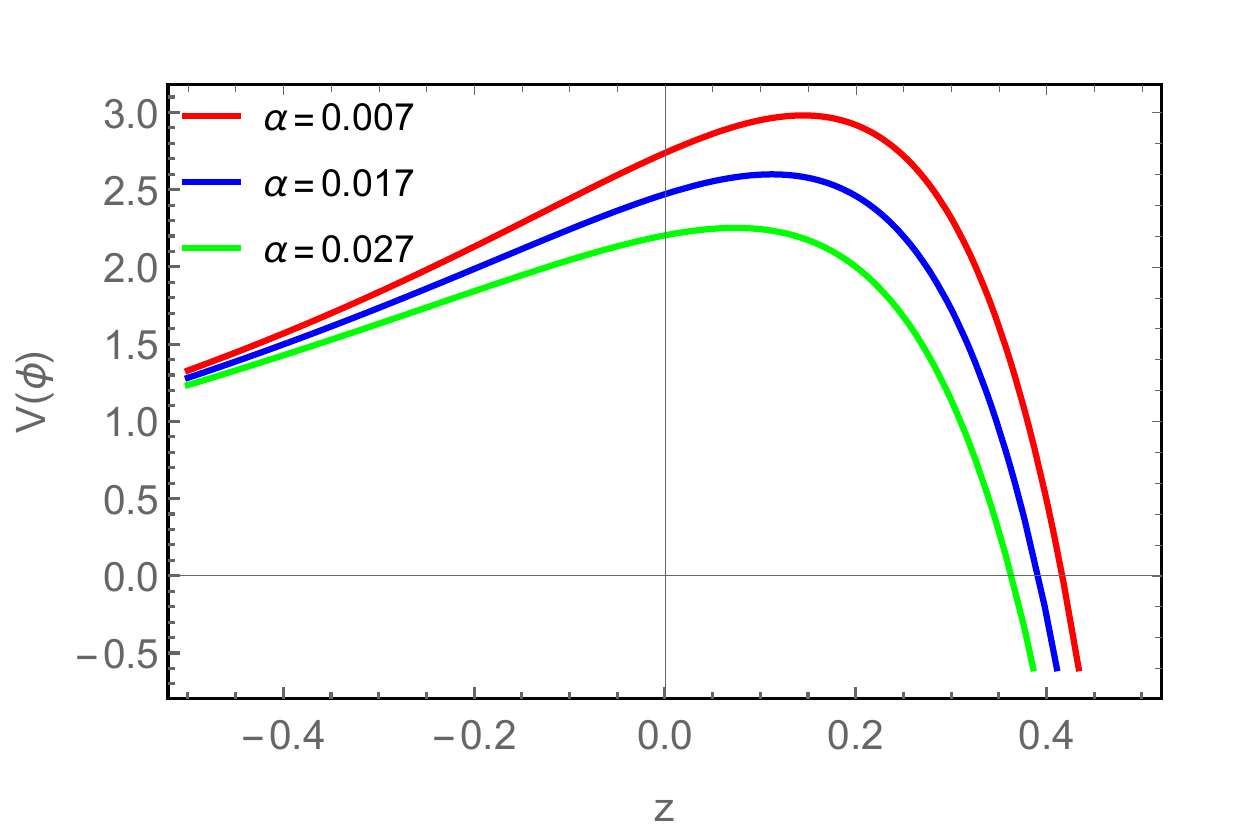}
\includegraphics[scale=0.50]{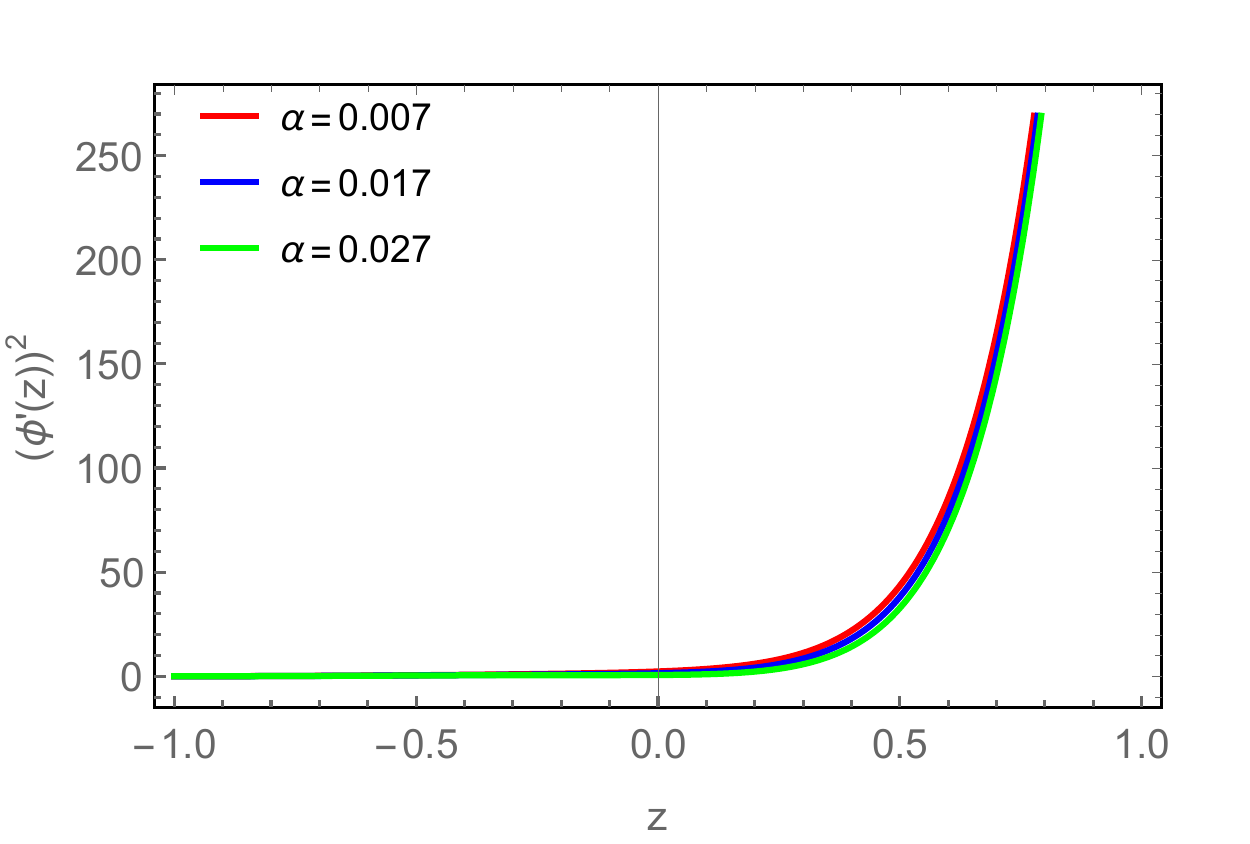}
\caption{Graphical behaviour of potential function (left panel) and squared slope of reconstructed scalar field (right panel) versus $z$ with the representative value of the parameters, $\eta=0.32$, $\nu=0.614$, $\beta=0.00018$. }
\label{Fig5}
\end{figure}

Apart from the cosmological constant, another important candidate for dark energy is the scalar field with a slowly varying potential. This can also be used for the cosmological mechanism such as the inflation, which can be constrained through Cosmic Microwave Background(CMB) observations. This is compatible with a number of modified theories
of gravity. The scalar field models can also be studied in the modified theory of gravity, where the effective energy momentum tensor with geometrical origin can be obtained \cite{Felice10,Sotiriou10}. Here we wish to reconstruct our model by introducing the scalar field. The cosmic acceleration phenomena can be modelled through the scalar field $\phi$ which can either be quintessence like or phantom like with the EoS parameter, ($\omega=p_{\phi}/\rho_{\phi}$) being $\omega \geq -1$ or $\omega \leq -1$ respectively. The action for the scalar field reconstruction is given by,
\begin{eqnarray}
S=\int d^4x \sqrt{-g} \left[\frac{R}{16 \pi} + \frac{\epsilon}{2} {\partial_\mu}{\phi}{\partial^\mu}{\phi}-V(\phi)\right] \label{20}
\end{eqnarray}
where $\epsilon = +1$ for quintessence field and $\epsilon = -1$ for phantom field, $V(\phi)$ is potential function of the scalar filed. In a flat Friedman background, the energy density and pressure are derived for the quintessence field as,
\begin{eqnarray}
\rho_{\phi}=\frac{1}{2} \left(\frac{d\phi}{dt} \right)^2 + V(\phi) \label{21}\\
p_{\phi}=\frac{1}{2} \left(\frac{d\phi}{dt} \right)^2 - V(\phi)\label{22}
\end{eqnarray}
From the above eqns. \eqref{21} and \eqref{22}, we can derive the scalar field and the potential function. The graphical behaviour of the potential function with respect to the redshift function can be observed in FIG. \ref{Fig5}(left panel). For the representative values of the parameter $\alpha$, the potential function evolving out from a infinite high value with a sharp increase and after attained the peak value at $z\simeq0.2$, it gradually decreases. At late times, the curve approaches to a small value. Another observation is that higher in value of $\alpha$ the curve is more steep. The squared slope of reconstructed scalar field approaching to zero at late times and there is no change in the behaviour with the varying value of $\alpha$ [FIG. \ref{Fig5}(right panel)].    

\subsection{Stability Analysis} 
The study of stability analysis in extended gravity cosmological model has become necessary, as several assumptions are considered to deal with the field equations. The degree of the generality of the assumptions made has become difficult to assess. So, the qualitative properties of the field equations are need to be analysed to strengthen the result. One of the approaches to perform this analysis is the stability analysis \cite{Wainright97,Charters01}. Here we wish to study the stability of the cosmological solutions of the $F(R,G)$ theory presented in this work under linear homogeneous and isotropic perturbations. For this purpose, we consider a pressureless dust FRW background whose general solution may be $H(t)=H_b(t)$. Here we consider perturbations of the Hubble parameter and the energy density around the arbitrary solutions $H_b(t)$ as \cite{de_la_cruz-dombriz/2012,Agrawal2021}
\begin{eqnarray}
H(t) &=& H_b\left(1+\delta (t)\right),\\
\rho (t) &=& \rho_b\left(1+\delta_m (t)\right),
\end{eqnarray}
where $\delta_m(t)$ and $\delta(t)$ are the respective deviations from the background energy density and the Hubble parameter. The functional $F(R,G)$ may be expanded around the solution $H(t)=H_b(t)$ as
\begin{equation}
F(R,G) = F(R_b,G_b)+(R-R_b)F_R(R=R_b)+(G-G_b)F_G(G=G_b)+\mathcal{O}^2,
\end{equation}
where $R_b$ and $G_b$ correspond to the background solutions. The respective first derivatives $F_R(R=R_b)$ and $F_G(G=G_b)$ are evaluated at $R_b$ and $G_b$. The term $\mathcal{O}^2$ includes all terms containing the higher powers in $R$ and $G$.

Using the perturbative approach in the equivalent FRW equation, it is possible to obtain an evolution equation for the linear homogeneous perturbations which may be written in the form
\begin{equation}
\kappa^2 \rho_b(t)\delta_m(t)=\chi_1(F_b,F^n_b)\ddot{\delta}(t)+\chi_2(F_b,F^n_b)\dot{\delta}(t)+\delta (t),\label{eq:26}
\end{equation}
where the coefficients $\chi_1(F_b,F^n_b)$ and $\chi_2(F_b,F^n_b)$ are functions of the functional $F(R,G)$ and its derivatives evaluated at the background.

From the continuity equation, we may get another evolution equation for the perturbations as
\begin{equation}
\dot{\delta}_m(t)+3H_b(t)\delta(t)=0.
\end{equation}

In this work, we have considered the functional $F(R,G)=R+\alpha R^2+\beta G^2$. Assuming that, GR should be recovered from the present model at some limit, we may neglect the contributions from the higher derivatives of the functional $F(R,G)$. In such a case, Eq. \eqref{eq:26} may be reduced to
\begin{equation}
\chi_3(t) \rho_b(t)\delta_m(t) = \ddot{\delta}(t)+\chi_4(t)\delta(t),
\end{equation} 
where 
\begin{eqnarray}
\chi_3(t) &=& \frac{\kappa^2}{18\gamma(t)},\\
\chi_4(t) &=& \frac{1}{3\gamma(t)}(1+2\alpha R_b),\\
\gamma(t) &=& 2\alpha+\beta H_b^4(t).
\end{eqnarray}

One may note that, in the GR limit, stability for the present model may be achieved provided we have $\gamma(t)>0$. In the present work, we have chosen the values of the coupling parameters $\alpha$ and $\beta$ to be positive so that the condition $\gamma(t)>0$ is satisfied through out to provide a model which may be stable under linear homogeneous and isotropic perturbations.

\section{Conclusion}
We have studied the late time cosmic acceleration issue in $F(R,G)$ theory of gravity in presence of time varying deceleration parameter. The model shows quintessence like behaviour and is dynamically stable under linear homogeneous and isotropic perturbations. The scale factor chosen here provides the deceleration parameter that becomes positive at early time and negative at late times. So, the other two parameters of the $F(R,G)$ function has been fixed in such a manner that the model shows an accelerating behaviour. Both the parameters that associated with the function $F(R,G)$ contributing to the behaviour of the functional as well the model. However, while relating to the accelerating behaviour, the parameter $\alpha$, associated with the Ricci scalar $R$ is more significant than the parameter $\beta$ associated with the Gauss-Bonnet invariant $G$. At the same time the varying $\beta$ does not show any significant changes in the behaviour whereas higher value of $\alpha$ shows some changes in the transition behaviour. The behaviour of the EoS parameter at late time becomes insensitive to the choice of $\alpha$ since all the trajectories of EoS parameter for different choices of $\alpha$ behave alike at late phase. Another important result we obtained on the geometrical parameters, the deceleration parameter approaches to $-1$ at late times, however the $(j,s)$ pair merging close to $(0.46,-1)$. Since the model favours quintessence behaviour, this value of $(j,s)$ pair is expected. The violation of SEC further validates the accelerating behaviour of the model in a modified theory of gravity. The important results of the present work are summarized in Tables II and III. Finally, we can say that the model presented here is another suitable extension of $f(R)$ gravity that provides quintessence like behaviour at late phase. Hence, this model can be another approach in the search of the geometrical alternative to dark energy phenomena.
\newpage
\begin{table}
\caption{Results for the EoS parameter and the cosmological parameters as estimated at the present epoch.}
\centering
\begin{tabular}{c|c|c|c}
\hline
\hline
Parameters  & $\alpha=0.007$  & $\alpha=0.017$ & $\alpha=0.027$ \\

\hline

$\omega$     &-0.40 &  -0.54& -0.74\\
$q$          &-0.13&-0.13&-0.13 \\
$j$	         &0.46&0.46&0.46\\
$s$          &-1&-1&-1\\
$l$          &4.7&4.7&4.7\\
$V(\phi)$    &2.73&2.47&2.20\\
\hline
\end{tabular}
\end{table}
\begin{table}
\caption{Important results of the present work.}
\centering
\begin{tabular}{c|c|c}
\hline
\hline
Tests & Early times ($z>>1$)& Late times ($z\simeq -1$) \\
\hline
energy conditions & &\\
DEC		& violated & satisfied\\
WEC     & satisfied & satisfied   \\
NEC     & satisfied & satisfied  \\
SEC		& satisfied & violated \\
\hline
Dynamical Stability     & stable & stable  \\
\hline
$Om(z)$     & shows &shows\\
& cosmological constant & cosmological constant\\
			&  behaviour            & behaviour\\
\hline
\end{tabular}
\end{table}

\section*{Acknowledgement} SVL acknowledges the financial support provided by University Grants Commission (UGC) through Junior Research Fellowship  (UGC Ref. No.:191620116597) to carry out the research work. BM and SKT thank IUCAA, Pune, India for providing support through the visiting Associateship program. The authors are thankful to P.H.R.S. Moraes, Universidade de Sao Paulo, Instituto de Astronomia, Sao Paulo, Brazil for his constructive suggestions during the preparation of the manuscript. The authors are thankful to the anonymous referees for their valuable comments and suggestions for the improvement of the paper.



\begin{thebibliography}{99}
\section*{References}

\bibitem{astier/2012} P. Astier and R. Pain, Compt. Rend. - Phys. {\bf 13}, 521 (2012).
\bibitem{bengaly/2020} C.A.P. Bengaly, Month. Not. Roy. Astron. Soc.: Lett. {\bf 499}, L6 (2020).
\bibitem{copeland/2006} E.J. Copeland et al., Int. J. Mod. Phys. D {\bf 15}, 1753 (2006).
\bibitem{frieman/2008} J.A. Frieman et al., Ann. Rev. Astron. Astrophys. {\bf 46}, 385 (2008).
\bibitem{li/2011} M. Li et al., Comm. Theor. Phys. {\bf 56}, 525 (2011).
\bibitem{riess/1998} A.G. Riess et al., Astron. J. {\bf 116}, 1009 (1998).
\bibitem{peebles/2003} P.J. Peebles and B. Ratra, Rev. Mod. Phys. {\bf 75}, 559 (2003).
\bibitem{carroll/2001} S.M. Carroll, Liv. Rev. Rel. {\bf 4}, 1 (2001).
\bibitem{weinberg/1989} S. Weinberg, Rev. Mod. Phys. {\bf 61}, 1 (1989).
\bibitem{martin/2012} J. Martin, Compt. Rend. - Phys. {\bf 13}, 566 (2012).
\bibitem{sotiriou/2010} T.P. Sotiriou and V. Faraoni, Rev. Mod. Phys. {\bf 82}, 451 (2010).
\bibitem{de_felice/2010} A. De Felice and S. Tsujikawa, Liv. Rev. Rel. {\bf 13}, 3 (2010).
\bibitem{nojiri/2011} S. Nojiri and S.D. Odintsov, Phys. Rep. {\bf 505}, 59 (2011).
\bibitem{matsumoto/2013} J. Matsumoto, Phys. Rev. D {\bf 87}, 104002 (2013).
\bibitem{carloni/2008} S. Carloni et al., Phys. Rev. D {\bf 77}, 024024 (2008).
\bibitem{Samanta19} G.C. Samanta and N. Godani, Eur. Phys. J. C, {\bf 79}, 623 (2019).

\bibitem{Samanta19a}G.C. Samanta and N Godani, Mod. Phys. Lett. A, {\bf 34}, 1950224 (2019).
 
\bibitem{Godani20} N. Godani and G.C. Samanta, Eur. Phys. J. C, {\bf 80}, 30 (2020).

\bibitem{Bohmer07} C.G. Bohmer, L. Hollenstein and F.S.N. Lobo, Phys. Rev. D, {\bf 76}, 084005 (2007).

\bibitem{Shah19} P. Shah, G.C. Samanta,Eur. Phys. J. C, {\bf 79}, 414 (2019). 

\bibitem{Pretel20} J.M.Z. Pretel et al., JCAP, {\bf 11 }, 048 (2020).

\bibitem{frolov/2008} A.V. Frolov, Phys. Rev. Lett. {\bf 101}, 061103 (2008).
\bibitem{kobayashi/2008} T. Kobayashi and K.-I. Maeda, Phys. Rev. D {\bf 78}, 064019 (2008).
\bibitem{kobayashi/2009} T. Kobayashi and K.-I. Maeda, Phys. Rev. D {\bf 79}, 024009 (2009).
\bibitem{goswami/2014} R. Goswami et al., Phys. Rev. D {\bf 90}, 084011 (2014).
\bibitem{chiba/2003} T. Chiba, Phys. Lett. B {\bf 575}, 1 (2003).
\bibitem{chiba/2007} T. Chiba et al., Phys. Rev. D {\bf 75}, 124014 (2007).
\bibitem{olmo/2007} G.J. Olmo, Phys. Rev. D {\bf 75}, 023511 (2007).
\bibitem{de_laurentis/2015} M. De Laurentis et al., Phys. Rev. D {\bf 91}, 083531 (2015).
\bibitem{wu/2015} B. Wu and B.-Q. Ma, Phys. Rev. D {\bf 92}, 044012 (2015).
\bibitem{santos_da_costa/2018} S. Santos da Costa et al., Class. Quant. Grav. {\bf 35}, 075013 (2018).
\bibitem{farasat_shamir/2018} M. Farasat Shamir and A. Komal, Comm. Theor. Phys. {\bf 70}, 190 (2018).
\bibitem{odintsov/2019} S.D. Odintsov et al., Nucl. Phys. B {\bf 938}, 935 (2019).
\bibitem{kumar_sanyal/2020} A. Kumar Sanyal and C. Sarkar, Class. Quant. Grav. {\bf 37}, 055010 (2020).
\bibitem{singh/2021} R. Singh, New Astron. {\bf 85}, 101513 (2021). 
\bibitem{de_la_cruz-dombriz/2012} \'A. de la Cruz-Dombriz and D. S\'aez-G\'omez, Class. Quant. Grav. {\bf 29}, 245014 (2012).
\bibitem{de_felice/2011} A. de Felice et al., Phys. Rev. D {\bf 83}, 104035 (2011).
\bibitem{elizalde/2010} E. Elizalde et al., Class. Quant. Grav. {\bf 27}, 095007 (2010).
\bibitem{lima/2008} M.P. Lima et al., Phys. Lett. B {\bf 668}, 83 (2008).
\bibitem{morris/1988} M.S. Morris et al., Phys. Rev. Lett. {\bf 61}, 1446 (1988).
\bibitem{ford/1995} L.H. Ford and T.A. Roman, Phys. Rev. D {\bf 51}, 4277 (1995).
\bibitem{tipler/1978} F.J. Tipler, Phys. Rev. D {\bf 17}, 2521 (1978).
\bibitem{mehdizadeh/2015} M.R. Mehdizadeh et al., Phys. Rev. D {\bf 91}, 084004 (2015).
\bibitem{pennalima/2018} M. Penna-Lima et al., Eur. Phys. J. C {\bf 79}, 175 (2019).


\bibitem{Mishra15} B.Mishra and S.K. Tripathy, Mod.Phys. Lett. A, {\bf 30}, 1550175 (2015).

\bibitem{SKT2020} S. K. Tripathy, S. K. Pradhan, Z. Naik, D. Behera, B. Mishra, Phys. of Dark Univ., {\bf 30}, 100722 (2020).

\bibitem{SKT2021} S. K. Tripathy, B. Mishra and S. Ray, to appear in Int. J. Mod. Phys. D (2021).

\bibitem{SKT21a} S.K. Tripathy, B. Mishra, M. Khlopov and S. Ray, arXiv:2106.04368.

\bibitem{Pati2021} L. Pati, B. Mishra and S. K. Tripathy, Phys. Scr., {\bf 96}, 105003  (2021).

\bibitem{Hawking73} S. Hawking and G.F.R. Ellis, The Large Scale Structure of Space-Time, Cambridge University Press, (1973). 

\bibitem{Poisson04} E. Poisson, A Relativist's Toolkit: The Mathematics of Black Hole Mechanics, Cambridge University Press, (2004).

\bibitem{Carroll04} S. Carroll,  Space-time and Geometry: An Introduction to General Relativity, Cambridge University Press, (2004).

\bibitem{Nojiri11} S. Nojiri and S.D. Odintsov, Phys. Rep., {\bf 505}, 59 (2011).

\bibitem{Capozziello11}S. Capozziello and M. De Laurentis, Phys. Rep., {\bf 509} (2011).

\bibitem{Weinberg72} S. Weinberg, Gravitation and Cosmology, John Wiley and Sons, New York (1972).

\bibitem{Riess16} A. G. Riess et al., Astrophys. J., {\bf 826}, 56 (2016).

\bibitem{Riess18} A. G. Riess et al., Astrophys. J., {\bf 861}, 126 (2018).

\bibitem{Aghanim20} N. Aghanim et al, Astron. Astrophys., {\bf 641}, A6 (2020). 

\bibitem{Alam03} U. Alam et al., Mon. Not. R. Astron. Soc., {\bf344}, 1057 (2003). 

\bibitem{Sahni08} V. Sahni et al., Phys. Rev. D, {\bf 78}, 103502 (2008).

\bibitem{Ding15} X. Ding et al., The Astrophysical Journal Letters, {\bf 803}, L22 (2015).

\bibitem{Zheng16} X. Zheng et al., The Astrophysical Journal, {\bf 825}, 17 (2016).

\bibitem{Qi18} Jing-Zhao Qi et al., Res. Astron. Astrophys., {\bf 18}, 066 (2018).

\bibitem{Yahya14} S. Yahya et al., Phys. Rev. D, {\bf 89}, 023503 (2014).

\bibitem{Qi16} Jing-Zhao Qi et al. arXiv:1606.00168 (2016).

\bibitem{Felice10} A. De Felice and S. Tsujikawa, Liv. Rev. Rel., {\bf13}, 3 (2010).
 
\bibitem{Sotiriou10} T.P. Sotiriou and V. Faraoni, Rev. Mod. Phys., {\bf 82}, 451(2010).

\bibitem{Wainright97} J. Wainright, G. Ellis, Dynamical systems in Cosmology, Cambridge University Press, (1997).

\bibitem{Charters01} T.C. Charters et al., Class. Quantum Gravt., {\bf 18}, 1703 (2001).

\bibitem{Agrawal2021} A. S. Agrawal, F. Tello-Ortiz, B. Mishra and S. K. Tripathy, arxiv: 2111.02894.

\bibitem{Balbi07} A. Balbi et al., Phys. Rev.D, {\bf 76}, 103519 (2007).

\bibitem{Xu12} L. Xu et al., Phys. Rev.D, {\bf 85}, 043003 (2012).

\bibitem{Tripathy20a} S. K. Tripathy et al., Phys. Scr., \textbf{95}, 115001 (2020).


\bibitem{Mishra21} B. Mishra, S.K. Tripathy, S. Tarai, J. Astrophys. Astron., {\bf 42}, 2 (2021).

\end{thebibliography}
\end{document}